\documentclass[ amsmath, amssymb, aps, pra, 12pt]{revtex4-2}

\usepackage{graphicx}
\usepackage{dcolumn}
\usepackage{bm}
\usepackage{hyperref}
\usepackage{xcolor}

\begin{document}

\title{Josephson-like oscillations in toroidal spinor Bose-Einstein condensates: \textit{a prospective symmetry probe}}

\author{Mário H. Figlioli Donato}
\email{mario.donato@usp.br}
\author{Sérgio R. Muniz}
\email{srmuniz@ifsc.usp.br}
\affiliation{
Instituto de Física de São Carlos, Universidade de São Paulo, IFSC-USP\\
Caixa Postal 369, CEP 13560-970, São Carlos, SP, Brazil.
}
\date{23/04/2022}

\begin{abstract}
Josephson junctions are essential ingredients in the superconducting circuits used in many existing quantum technologies. Additionally, ultracold atomic quantum gases have also become essential platforms to study superfluidity. Here, we explore the analogy between superconductivity and superfluidity to present an intriguing effect caused by a thin finite barrier in a quasi-one-dimensional toroidal spinor Bose--Einstein condensate (BEC). In this system, the atomic current density flowing through the edges of the barrier oscillates, such as the electrical current through a Josephson junction in a superconductor, but in our case, there is no current circulation through the barrier.  We also show how the nontrivial broken-symmetry states of spinor BECs change the structure of this Josephson-like current, creating the possibility to probe the spinor symmetry, solely using  measurements of this superfluid current. \\
\\ Published in \textit{Symmetry} \textbf{2022}, 14(5), 867 -- https://doi.org/10.3390/sym14050867
\end{abstract}

\maketitle

\newpage
\section{Introduction} \label{S1}

The Josephson effect \cite{Josephson1962,Josephson1974} is one of the most relevant phenomena in \textit{superconductivity}; for his theoretical predictions, Brian D. Josephson received the Nobel Prize in Physics in 1973. In a superconductor, the phenomenon is characterized by the tunneling of \textit{Cooper pairs} through a junction, or a weak link, represented by a potential energy barrier between two superconductor regions \cite{Barone-Paterno,tinkham,supercond}. When a voltage $V$ is applied to these regions, the electrical current in the junction, $I_j$, oscillates according to the expression: 
\begin{equation}\label{JosephsonCurrent}
    I_j = I_{0} \,\text{sin}(\phi);\quad\text{ with } \frac{d\phi}{dt} = \frac{2eV}{\hbar},
\end{equation}
where $I_{0}$ is the current amplitude and $\phi$ is the angle representing the phase difference of the order parameter between each superconducting region \cite{tinkham}. There is a conceptual analogy between superconductivity and superfluidity \cite{Analogia}, given by the lack of resistance in the flow of the electrical current, or, equivalently, the lack of viscosity in the flow of a superfluid. In fact, since superfluidity emerges naturally in Bose--Einstein condensates (BECs) produced in dilute ultracold Bose gases \cite{pethick,Pitaevskii2016,Kawaguchi_Ueda,SKurn-Ueda2013,Yukalov2018,penrose-onsager,Bogoliubov,Zapata1998,Williams1999}, the analogue of the Josephson effect has already been studied in experiments \cite{Hall1998b,Albiez2005,Levy2007,SQUID}.

For \textit{scalar} BECs \cite{pethick,Pitaevskii2016} (i.e., condensates with a scalar order parameter: \mbox{$\psi=\sqrt{n}e^{i\theta}$}), the superfluid velocity is always proportional to the global phase gradient ($\nabla\theta$), and it represents the complete \textit{structural symmetry} of the system. On the other hand, for \textit{spinor} BECs \cite{Kawaguchi_Ueda,SKurn-Ueda2013,Yukalov2018} (i.e., condensates with multi-component order parameter: 
 $[\boldsymbol{\psi}]_m = \psi_m$), the superfluid flow is related to the properties of their symmetry and topology, which can generate nontrivial spatial spin textures and current distributions not found in scalar BEC systems. In fact, there are many topological states in spinor BECs analogous to particles and structures studied across several areas of physics, such as \textit{skyrmions}, \textit{Dirac monopoles}, \textit{knot solitons}, \textit{vortices}, \textit{half-vortices}  \cite{Kawaguchi_Ueda,polarcore,carga_artigo,Nos_condensados, Dirac,skyrmions}, to name a few. 

\begin{figure}
\centering   
\hspace{-0.4CM}\includegraphics[width=7 cm]{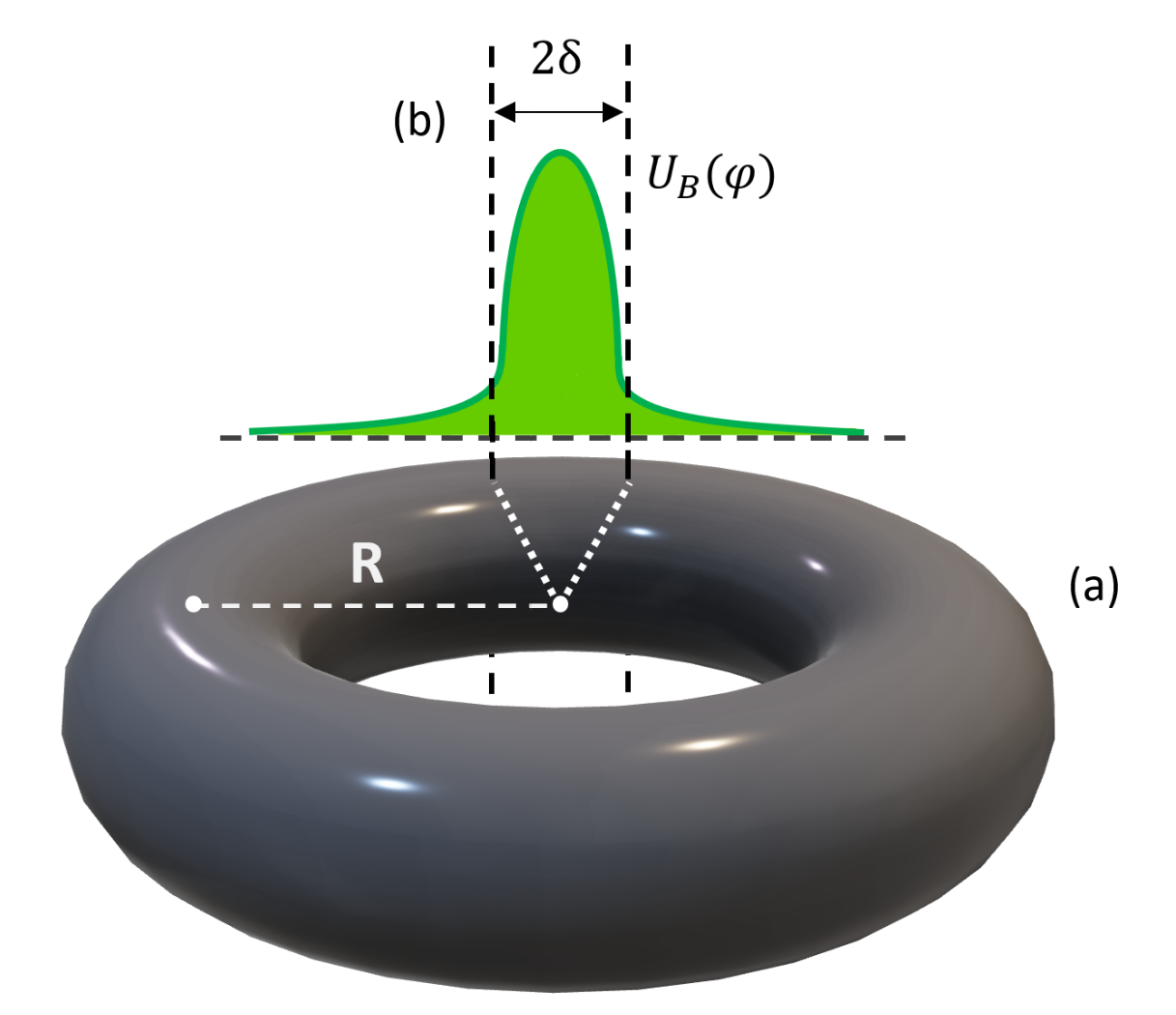}    
\caption{Schematic illustration of the external potentials applied to the condensate: (\textbf{a}) effective toroidal region $\mathcal{T}$ (with average radius $R$) where the potential $U_{\text{trap}}$ confines the condensate; (\textbf{b})~repulsive barrier $U_B = U_B(\varphi)$, for $\varphi$ the azimuthal angle in cylindrical coordinates $(r,\varphi,z)$. Here, we consider thin angular widths $2\delta\ll 2\pi$.} \label{Fig1}
\end{figure}

Here, based on the analogy between superfluidity and superconductivity, we study a Josephson-like oscillating current produced in toroidal spinor Bose--Einstein condensates due to the presence of a thin finite energy barrier (Figure \ref{Fig1}), and we show that the current density at the edges of the barrier behaves similarly to $I_j$, and depends on the structural symmetry of the spinor condensate. 

\section{General Modeling of Spinor BECs} \label{S2}

For a system with atomic spin $f=1$, the order parameter of a spinor BEC satisfies the following vector Gross--Pitaevskii equation (GPE) \cite{Kawaguchi_Ueda}:
\begin{equation}\label{GPE}
    i\hbar\,\frac{\partial}{\partial t}\boldsymbol{\psi} = \left[-\frac{\hbar^2 \nabla^2}{2M} + U_{\text{total}} - p\mathbf{f}_z + q\mathbf{f}_z^2 + c_0 n +c_1\vec{F}\cdot\vec{\mathbf{f}}\,\right]\boldsymbol{\psi} \,.
\end{equation}

In the equation, the bold fonts indicate matrix quantities related to the vector order parameter $\boldsymbol{\psi}=(\psi_{f}(\vec{r},t),...,\psi_{-f}(\vec{r},t))^T$, and where $n = \boldsymbol{\psi}^{\dagger}\boldsymbol{\psi}=\sum_m{|\psi_m(\vec{r},t)|^2}=\sum_m{n_m(\vec{r},t)}$ is the number density, with spin components $\psi_m(\vec{r},t)=\sqrt{n_m(\vec{r},t)}\,e^{i\theta_m(\vec{r},t)}$, $\vec{F} = \boldsymbol{\psi}^{\dagger}\vec{\mathbf{f}}\,\boldsymbol{\psi}$ is the spin density, and $\vec{\mathbf{f}} = (\mathbf{f}_x,\mathbf{f}_y,\mathbf{f}_z)$ is the spin operator (in the $f=1$ representation). The constants $p\propto |\vec{B}|$ and $q\propto |\vec{B}|^2$ are related to the linear and quadratic Zeeman effects, respectively, when an external magnetic field, $\vec{B}$, is applied. The system's nonlinear interaction ``strength'' is modeled by the coefficients $c_0$ and $c_1$ (see Reference \cite{Kawaguchi_Ueda} for details). 

In this work, we consider a toroidal trapping potential, $U_{\text{trap}}$, with a repulsive barrier, $U_B$, restricted to a region of small azimuthal angle. Therefore, the total external potential is given by $U_{\text{total}} = U_{\text{trap}} + U_B$, as sketched in Figure \ref{Fig1}. 
We assume that $U_{\text{total}}$ provides the necessary means to trap all the components of the spinor BEC in the torus. For instance, by using the appropriate optical trapping techniques~\cite{Grimm2000}. Optical traps have the advantage of providing flexible control over the potential symmetry when combined, for example, with digital holography \cite{Pasienski2008,Holografia}, direct imaging \cite{Gauthier2016}, or phase-mapping \cite{Faleiros2021} of sharp optical patterns, to design almost arbitrary shapes. In fact, a toroidal optical potential with a controllable energy barrier similar to the one described here has been demonstrated in~\mbox{\cite{Ramanathan2011,Wright2013,Eckel2014}}, and a toroidal potential with spinor BEC in \cite{Beattie2013}.

The order parameter of a spinor BEC, represented as a column matrix, can always be factored into $\boldsymbol{\psi} = \psi (\vec{r},t) \,\boldsymbol{\zeta}$, as the product of a \textit{scalar part}, $\psi (\vec{r},t) = \sqrt{n(\vec{r},t)} \, e^{i \theta (\vec{r},t)}$, and a~normalized \textit{spinor part}, $\boldsymbol{\zeta}$ (with $\boldsymbol{\zeta}^{\dagger}\boldsymbol{\zeta} = 1$) \cite{Kawaguchi_Ueda}. This vector description has a gauge symmetry with \textit{artificial gauge fields} related to $\boldsymbol{\zeta}$. These fields emerge naturally from the description, and are analogous to the scalar and vector electromagnetic potentials \cite{Kawaguchi_Ueda, Revspielman}:
\begin{equation}
    \Phi = -i\boldsymbol{\zeta}^{\dagger}\frac{\partial}{\partial t}\boldsymbol{\zeta},
\end{equation}
\begin{equation}
    \Vec{\mathcal{A}} = i\boldsymbol{\zeta}^{\dagger}\Vec{\nabla}\boldsymbol{\zeta}.
\end{equation}

In particular, the superfluid number current density, $\vec{J} = n\,\vec{v}$, depends linearly on the vector potential $\vec{\mathcal{A}}$, as the superfluid velocity $\vec{v}$ of a spinor BEC is given by:
\begin{equation}\label{campo de velocidades}
    \vec{v} = \frac{\hbar}{M}(\vec{\nabla} \theta -\vec{\mathcal{A}}).
\end{equation}

Note that the action of the artificial vector potential is analogous to the electromagnetic vector potential on the expression of the electrical current in a superconductor \cite{tinkham}.
In addition, some of the topological properties of the condensate are determined by the \textit{circulation} of $\vec{v}$ (i.e., $\oint\vec{v}\cdot \vec{dl}$), and, therefore, they may depend on \textit{artificial magnetic fluxes}, given by a synthetic magnetic field $\vec{\mathcal{B}} \equiv \hbar \vec{\nabla}\times\vec{\mathcal{A}}$ \cite{Kawaguchi_Ueda}.

In general, the current density of a spinor BEC is defined in the usual manner, replacing the scalar order parameter by its vector form
\begin{equation}
     \vec{J}= \frac{\hbar}{M} \left( \boldsymbol{\psi}^{\dagger} \vec{\nabla} \boldsymbol{\psi} - (\vec{\nabla} \boldsymbol{\psi}^{\dagger}) \boldsymbol{\psi} \right) = \frac{\hbar}{M} \text{Im} \{ \boldsymbol{\psi}^{\dagger} \vec{\nabla} \boldsymbol{\psi} \}, 
\end{equation}
therefore,
\begin{eqnarray}
    \vec{J} &=& \frac{\hbar}{M} \sum_{m=-f}^{f} \text{Im} \{ \psi_m^{*}(\vec{r},t) \vec{\nabla} \psi_m(\vec{r},t) \} \\
    &=& \sum_{m=-f}^{f} \vec{J}_m(\vec{r},t)
\end{eqnarray}

Alternatively, using the spin components $\psi_m(\vec{r},t)$, one could write
\begin{equation}
    \vec{J}_m = \frac{\hbar}{M} n_m(\vec{r},t) \vec{\nabla} \theta_m (\vec{r},t) = n_m(\vec{r},t) \, \vec{v}_m (\vec{r},t), 
\end{equation}
where the velocity spin components are $\vec{v}_m(\vec{r},t)=\frac{\hbar}{M}\vec{\nabla} \theta_m (\vec{r},t)$.

\section{Modeling the Potential Barrier and Defining Some Approximations} \label{S3}

This section explains in detail how we define the potential barrier, $U_B$, and its general properties. Because we will use the quasi-one-dimensional (\textit{quasi-1D}) limit for the geometry of the condensate later, we consider the barrier $U_B(\varphi)$ is written in terms of a normalized angular distribution $f(\varphi)$, not depending on other spatial variables ($r,z$):
\begin{equation}
    U_B(\varphi) = U_0\,f(\varphi), \quad\text{ with } \int_{-\pi}^{\pi}f(\varphi)d\varphi = 1.
\end{equation}

We consider that the barrier is effectively restricted to an angular range $\varphi\in[-\delta,\delta]$, such that:
\begin{equation}
    \int_{-\pi}^{\pi}f(\varphi)d\varphi \approx \int_{-\delta}^{\delta}f(\varphi)d\varphi \approx 1.
\end{equation}

Integrating the GPE (\ref{GPE}) along the arc $\varphi\in[-\delta,\delta]$, with fixed $(r,z)$ in the thin barrier condition ($\delta \ll \pi$), leads to two dominant terms: 
\begin{equation}\label{eq8}
    \frac{\hbar^2}{2Mr^2} \left[\partial_{\varphi}\boldsymbol{\psi}(\delta) - \partial_{\varphi}\boldsymbol{\psi}(-\delta)\right]  \approx \int_{-\delta}^{\delta}U_B(\varphi)\boldsymbol{\psi}(\varphi)\,d\varphi,
\end{equation}
where the left-hand side follows directly from the fundamental theorem of calculus, and it is related to the momentum along the $\hat{\varphi}$ direction ($p_{\varphi}\hat{\varphi} = -\hat{\varphi}(i\hbar\partial_{\varphi})/r $). Note that we simplified the notation $\boldsymbol{\psi}(r,\varphi,z)\rightarrow \boldsymbol{\psi}(\varphi)$, given that we consider ($r,z$) fixed parameters. 

The remaining terms not computed in the last equation were neglected because for small $\delta$ they are proportional to $\delta$ (i.e., these terms are $\mathcal{O}(\delta)$ while the dominant terms are proportional to the unit), and can be neglected as one takes the limit of $\delta \rightarrow 0$.

Additionally, in the limit of small $\delta$, the integration of the barrier potential resembles the integration of a Dirac's delta (normalized) function, such that 
\begin{equation} \label{eq9}
    \lim_{\delta \rightarrow 0} \, \left[ \int_{-\delta}^{\delta}U_B(\varphi)\boldsymbol{\psi}\,d\varphi \right] =  U_0  \, \lim_{\delta \rightarrow 0} \, \left[  \int_{-\delta}^{\delta} f(\varphi) \boldsymbol{\psi}\,d\varphi \right] \approx  U_0 \, \boldsymbol{\psi}(0).
\end{equation}

Here, it is important to emphasize that, in practice, the derivative of the potential barrier must not exceed a limit given by the Landau critical velocity, otherwise, the flow is dissipated and the system loses its superfluid properties \cite{Ramanathan2011,velocidadeLandau}. To avoid such situations, one may always choose a larger radius R (see Figure \ref{Fig1}) so that the barrier $U_B(\varphi)$ is spatially ``smooth'' enough. We discuss this condition in Section  \ref{S6}.

Finally, we consider one last approximation. If the number of atoms in the condensate is kept the same, it is expected that the average cross-section area of $\mathcal{T}$ (at fixed $\varphi$) becomes smaller when $R$ is larger. So, for $R$ big enough, the condensate would be trapped in small $\Delta r$ and $\Delta z$ ranges, and the number density would become approximately a function of only one variable (\textit{quasi-1D approximation}):
\begin{equation}
    n(r,\varphi,z) \rightarrow \left\{\begin{matrix} \,n(\varphi),\quad\text{ if $(r,\varphi,z)\in\mathcal{T}$}; \\ 0, \quad\text{ otherwise}.\end{matrix} \right.
\end{equation}

Moreover, these conditions also restrict the direction of $\vec{J}$, such that:
\begin{equation}\label{eq11}
    \vec{J}\approx J_{\varphi}\hat{\varphi} = \frac{\hbar}{M r}\text{Im}\{\boldsymbol{\psi}^{\dagger}\partial_{\varphi}\boldsymbol{\psi}\}\hat{\varphi}.
\end{equation}

In the next section, we discuss how the barrier potential, $U_B$, acts on the current density and the consequences of choosing an order parameter with \textit{defined parity}.

\section{Current Density and the Parity of the Order Parameter} \label{S4}

Based on the discussion in the last section, we expect that, for relatively large $R$, a~toroidal condensate under the presented conditions, will behave similar to a quasi-1D BEC with the current density effectively restricted to the $\hat{\varphi}$ direction. Now, we are interested in finding an expression for the current density $J_{\varphi}(\delta)$ at the edges of the barrier $U_B$ (i.e., at $\varphi = \pm \delta$). To help us find such an expression, we define the following quantity:
\begin{equation}\label{eq12}
    \Delta J(\delta) := J_{\varphi}(\delta) - J_{\varphi}(-\delta).
\end{equation}

{Its}  interpretation is quite simple: if we multiply it by the area, $A_c$, of the cross-section of the condensate, it returns the rate of change in the number of atoms $N_B$ located inside the barrier region: 
\begin{equation}\label{continuityequation}
    \Delta J(\delta)\cdot A_c = - \frac{d}{dt}N_B.
\end{equation}

Using (\ref{eq11}) and (\ref{eq12}), one finds the following expression for $\Delta J(\delta)$:
\begin{eqnarray}\label{DeltaJ}
    \Delta J(\delta) \approx \frac{\hbar}{MR}\text{Im}\left\{\boldsymbol{\psi}^{\dagger}(\delta)\partial_{\varphi}\boldsymbol{\psi}(\delta)-\boldsymbol{\psi}^{\dagger}(-\delta)\partial_{\varphi}\boldsymbol{\psi}(-\delta) \right\}.
\end{eqnarray} 

To simplify it, we use symmetry arguments observing the \textit{parity} of the order parameter $\boldsymbol{\psi}$. It is noticeable that, in general, the GPE (\ref{GPE}) admits both \textit{even} and \textit{odd} solutions (with respect to the variable $\varphi$), and if we choose an even distribution $f(-\varphi) = f(\varphi)$, the order parameter follows
\begin{equation}
    \left \{ \begin{matrix} 
    \boldsymbol{\psi}(-\varphi)&=&\boldsymbol{\psi}(\varphi) &\rightarrow &\text{ even, }\\ \boldsymbol{\psi}(-\varphi)&=&-\boldsymbol{\psi}(\varphi) &\rightarrow &\text{ odd.}
    \end{matrix}\right.
\end{equation}

{Moreover,}  it is easy to show that $J_\varphi$ is always an \textit{odd} function if the parity of $\boldsymbol{\psi}$ is well defined. This way, it is clear that $\Delta J(\delta)$ allows us to calculate $J_{\varphi}(\delta)$ directly:
\begin{equation}
    \Delta J(\delta) = 2 J_{\varphi}(\delta).
\end{equation}

{In}  particular, if we choose $\boldsymbol{\psi}(\varphi)$ to be even, one finds that:
\begin{equation}
    \boldsymbol{\psi}^{\dagger}(\delta)\partial_{\varphi}\boldsymbol{\psi}(\delta)-\boldsymbol{\psi}^{\dagger}(-\delta)\partial_{\varphi}\boldsymbol{\psi}(-\delta) =
    \boldsymbol{\psi}^{\dagger}(\delta) \left[\partial_{\varphi}\boldsymbol{\psi}(\delta) - \partial_{\varphi}\boldsymbol{\psi}(-\delta)\right],
\end{equation}

\noindent and, using the Equations (\ref{eq8}) and (\ref{eq9}), we finally obtain:
\begin{equation}\label{RAWcurrent}
    J_{\varphi}(\delta) \approx \frac{U_0 R}{\hbar}\text{Im}\{\boldsymbol{\psi}^{\dagger}(\delta)\boldsymbol{\psi}(0)\}.
\end{equation}

Note that making $\boldsymbol{\psi}$ an even function is not the only way of finding such an approximation for the current difference $\Delta J(\delta)$. For instance, choosing $RU_0$ big enough in the complete expression for $\Delta J(\delta)$,
\begin{eqnarray}\label{DeltaJ_completa}
    \Delta J(\delta) \approx \frac{\hbar}{MR}\text{Im}\left\{\left[\boldsymbol{\psi}^{\dagger}(\delta)-\boldsymbol{\psi}^{\dagger}(-\delta)\right]\partial_{\varphi}\boldsymbol{\psi}(\delta) + \frac{2MR^2}{\hbar^2}U_0\boldsymbol{\psi}^{\dagger}(-\delta)\boldsymbol{\psi}(0)\right\},
\end{eqnarray}
should effectively return the same $\Delta J(\delta)$ value from the even case, regardless of whether the parity of $\boldsymbol{\psi}$ is defined or not.
However, choosing the even parity case for $\boldsymbol{\psi}$ leads to a symmetrical constraint of the BEC global phase, as we show in the next section (see Equations (\ref{circ1}) and (\ref{circ2})).

Recalling what we mentioned before, it is always possible to factorize the order parameter $\boldsymbol{\psi} = \psi\boldsymbol{\zeta}$, and the scalar term can be written as $\psi = \sqrt{n}\,e^{i\theta}$. Applying such factorization in (\ref{RAWcurrent}), we find:
\begin{equation}\label{RAREcurrent}
    J_{\varphi}(\delta) \approx \frac{U_0 R}{\hbar}\sqrt{n(\delta)n(0)}\,\text{Im}\left\{\boldsymbol{\zeta}^{\dagger}(\delta)\boldsymbol{\zeta}(0)e^{i[\theta(0)-\theta(\delta)]}\right\}.
\end{equation}

{According} to reference \cite{Kawaguchi_Ueda}, $\boldsymbol{\zeta}$ is highly dependent on the symmetrical and topological properties of the condensate. Therefore, we expect that $J_{\varphi}$ will also depend on such properties. Nevertheless, for now, to keep the analogy between the Josephson current (\ref{JosephsonCurrent}) and $J_{\varphi}(\delta)$ as simple as possible, we consider the case of a scalar BEC (i.e., $\boldsymbol{\zeta} \equiv 1$): 
\begin{equation}
    J_{\varphi}(\delta) \approx \left\{\frac{U_0 R}{\hbar}\sqrt{n(\delta)n(0)}\right\}\times \text{sin}[\theta(0)-\theta(\delta)]
\end{equation}

{Note}  that the current density $J_{\varphi}$ at the edges of the barrier ($\varphi=\pm\delta$) is proportional to the \textit{sine} of the phase difference between the \textit{barrier region} ($\varphi \sim 0$) and the \textit{remaining region} ($\varphi \gtrsim \delta$ or $\varphi \lesssim -\delta$), and the analogy to the Josephson current is (mathematically) clear. 

It is also possible to derive the time-evolution equation for the global phase difference $\Delta\theta = \theta(0)-\theta(\delta)$, similarly to the second equation in (\ref{JosephsonCurrent}). Here, we use the following relation between $\partial_t\theta$ and the local average energy per particle $\delta E/\delta n(\vec{r})$~\cite{pethick}:
\begin{equation}
    -\hbar \frac{\partial}{\partial t} \theta = \frac{\delta E}{\delta n(\vec{r})},
\end{equation}

\noindent where $\delta E/\delta n(\vec{r})$ is the functional derivative of the average energy of the condensate, $E$,  with respect to the number density~\cite{pethick}, $n(\vec{r})$, 
\begin{equation}\label{energy}
    E =\int \psi^{*} \left[-\frac{\hbar^2 \nabla^2}{2M} + U_{\text{total}} \right]\psi + \frac{c_0 n^2}{2}  \,d^3r\,.
\end{equation}

{If} we consider the phase difference $\Delta\theta = \theta(0)-\theta(\delta)$,
\begin{equation}\label{faseanalogia}
    \frac{\partial}{\partial t} [\Delta\theta] = \left. \frac{1}{\hbar}\frac{\delta E}{\delta n}\right|^{\varphi=\delta}_{\varphi=0},
\end{equation}
\noindent we find that the difference of the local energy per particle (right-hand side of (\ref{faseanalogia})) has the same role to $J_{\varphi}(\delta)$ as $2eV$ has to $I_j$ in Equation (\ref{JosephsonCurrent}).

\section{ The Broken-Symmetry Spinor BEC Case} \label{S5}

This section discusses the current $J_{\varphi}(\delta)$ beyond the scalar BEC case. For that, it is convenient to define the function $C(\delta)$ as the quantity that characterizes such current according to the spinor nature of the condensate:
\begin{equation}\label{B}
    C(\delta) := \text{Im}\left\{\boldsymbol{\zeta}^{\dagger}(\delta)\boldsymbol{\zeta}(0)e^{i[\theta(0)-\theta(\delta)]}\right\}.
\end{equation}

{Note}  that, in general,  
\begin{equation}
    C(\delta) \neq \text{sin} ( \Delta\theta ),
\end{equation}

\noindent because the \textit{spinor} part $\boldsymbol{\zeta}$ may not be trivial (i.e., $\boldsymbol{\zeta} \not\equiv 1$). Before discussing specific cases, we recall some useful properties of spinor BECs applied to the system analyzed here. 

Firstly, due to the defined parity of $\boldsymbol{\psi}$, it is simple to show that the superfluid velocity in the $\hat{\varphi}$ direction ($v_{\varphi}$) is an odd function. Therefore, its circulation must always be zero and the \textit{global phase circulation} is proportional to the \textit{artificial magnetic flux}:
\begin{equation}\label{circ1}
    \oint d\theta = \oint \vec{\mathcal{A}} \cdot \vec{dl} = \frac{1}{\hbar}\iint \vec{\mathcal{B}}\cdot \vec{dS},
\end{equation}

\noindent which, in several situations \cite{Kawaguchi_Ueda}, might be proportional to integer multiples of $2\pi$. 

In addition, the time-evolution equation for $\Delta\theta$ in the spinor BEC case is given by:
\begin{equation}
    \frac{\partial}{\partial t} [\Delta\theta] = \left.\left(\frac{1}{\hbar}\frac{\delta E}{\delta n} + \Phi\right)\right|^{\varphi=\delta}_{\varphi=0}\,\,,
\end{equation}
where the average energy $E$ is (for atomic spin $f=1$):
\begin{equation}\label{energy1}
    E =\int \boldsymbol{\psi}^{\dagger} \left[-\frac{\hbar^2 \nabla^2}{2M} + U_{\text{total}} - p\mathbf{f}_z + q\mathbf{f}_z^2\right]\boldsymbol{\psi} + \frac{c_0 n^2}{2} +\frac{c_1\vec{F}^2}{2} \,d^3r \,,
\end{equation}

\noindent and it implicitly depends on the gauge field $\vec{\mathcal{A}}$.

Moreover, according to reference \cite{Kawaguchi_Ueda}, spinor BECs are naturally described by \textit{broken-symmetry} states in the long-wavelength limit (i.e, when the characteristic dimensions of the condensate are much larger than its \textit{healing length}). 
Furthermore, the spinor part of these states is characterized by the following type of expansion:
\begin{equation}
    \boldsymbol{\zeta} = e^{-i\alpha\mathbf{f}_z}e^{-i\beta\mathbf{f}_y}e^{-i\gamma\mathbf{f}_z} \boldsymbol{\zeta}_0,
\end{equation}

\noindent where the parameters $(\alpha,\beta,\gamma)$ are functions of time and space, and represent an arbitrary unitary transformation $\cong SO(3)$ (i.e., they are Euler angles). For $f=1$ condensates, such transformation is equivalent to the following matrix:
\begin{equation}\label{matriz}
    e^{-i\alpha\mathbf{f}_z}e^{-i\beta\mathbf{f}_y}e^{-i\gamma\mathbf{f}_z} = \begin{pmatrix}e^{-i(\alpha+\gamma)}\text{cos}^2\frac{\beta}{2} & -\frac{e^{-i\alpha}}{\sqrt{2}}\text{sin}\beta & e^{-i(\alpha-\gamma)}\text{sin}^2\frac{\beta}{2}\\ \frac{e^{-i\gamma}}{\sqrt{2}}\text{sin}\beta & \text{cos}\beta& -\frac{e^{i\gamma}}{\sqrt{2}}\text{sin}\beta\\ e^{i(\alpha-\gamma)}\text{sin}^2\frac{\beta}{2} & \frac{e^{i\alpha}}{\sqrt{2}}\text{sin}\beta& e^{i(\alpha+\gamma)}\text{cos}^2\frac{\beta}{2}\end{pmatrix}.
\end{equation}

{The} \textit{constant generator} of the spinor part $\boldsymbol{\zeta}_0$ is directly related to the order parameter of a uniform condensate, whose possible states (ferromagnetic, polar, anti-ferromagnetic, and others) have been studied and classified \cite{Kawaguchi_Ueda,SKurn-Ueda2013,diagramadefases}. For generators with average spin in the $\hat{z}$ direction
($\boldsymbol{\zeta}_0^{\dagger}\vec{\mathbf{f}}\,\boldsymbol{\zeta}_0 = \boldsymbol{\zeta}_0^{\dagger}\mathbf{f}_z\boldsymbol{\zeta}_0\hat{z} = f_0\hat{z}$), the artificial electromagnetic potentials are \cite{Kawaguchi_Ueda}:
\begin{equation}\label{Aexpans}
    \vec{\mathcal{A}} = f_0\left( \text{cos}(\beta)\vec{\nabla}\alpha + \vec{\nabla}\gamma\right),
\end{equation}
\begin{equation}\label{PHIexpans}
    \Phi = -f_0\left( \text{cos}(\beta)\frac{\partial}{\partial t}\alpha + \frac{\partial}{\partial t}\gamma\right).
\end{equation}

So, we can rewrite (\ref{circ1}) as: 

\begin{equation}\label{circ2}
    \oint d\theta =f_0 \oint (\text{cos}(\beta)d\alpha + d\gamma),
\end{equation}
which is a constrain between $\theta$ and the Euler angles (given that the parity of $\boldsymbol{\psi}$ is defined).

Now, we will show two examples of how such broken-symmetry states affect the current $J_{\varphi}(\delta)$ and how it differs from the scalar BEC case. 

\subsection{Ferromagnetic States}

There are two families of \textit{ferromagnetic} states in spinor ($f=1$) BECs, called \textit{positive} and \textit{negative}, generated by $\boldsymbol{\zeta}_0^{\text{ferro}+} = (1,0,0)^T$ and $\,\boldsymbol{\zeta}_0^{\text{ferro}-} = (0,0, 1)^T$, respectively. Using (\ref{matriz}), one derives \cite{Kawaguchi_Ueda} the following spinor parts of the ferromagnetic states $\boldsymbol{\zeta}_{\text{ferro}\pm}$: 
\begin{equation}\label{ferro+}
    \boldsymbol{\zeta}_{\text{ferro}+} = e^{-i\gamma}\begin{pmatrix}e^{-i\alpha}\text{cos}^2\frac{\beta}{2} \\ \frac{1}{\sqrt{2}}\text{sin}\beta\\ e^{i\alpha}\text{sin}^2\frac{\beta}{2}\end{pmatrix} \text{ and  }\,\,\boldsymbol{\zeta}_{\text{ferro}-} = e^{i\gamma}\begin{pmatrix}e^{-i\alpha}\text{sin}^2\frac{\beta}{2} \\ \frac{-1}{\sqrt{2}}\text{sin}\beta\\ e^{i\alpha}\text{cos}^2\frac{\beta}{2}\end{pmatrix},
\end{equation}

\noindent from which one can compute all the \textit{symmetrical} and \textit{topological} properties. With these expressions, we directly find how the parameters ($\alpha,\beta,\gamma$) act on $C(\delta)$:

\begin{eqnarray}
    C_{\text{ferro}\pm}(\delta) &=& \text{cos}^2\frac{\beta(0)}{2}\text{cos}^2\frac{\beta(\delta)}{2}\text{sin}[\Delta\theta\mp(\Delta\gamma+\Delta\alpha)] + \frac{1}{2}\text{sin}\beta(0)\text{sin}\beta(\delta)\text{sin}[\Delta\theta\mp\Delta\gamma] \nonumber\\ &+& \text{sin}^2\frac{\beta(0)}{2}\text{sin}^2\frac{\beta(\delta)}{2}\text{sin}[\Delta\theta\mp(\Delta\gamma-\Delta\alpha)],
\end{eqnarray}

\noindent for $\Delta G := G(0)-G(\delta)$, when $G = \theta, \alpha$ or $\gamma$. 

Note that $C_{\text{ferro}\pm}(\delta)$ is not equal to $\mathrm{sin}[\Delta\theta]$ in general. This shows that the current $J_{\varphi}(\delta)$ in BECs is sensitive to their nontrivial structural spinor properties. 
However, we are still able to access the scalar limit, because $C_{\text{ferro}\pm}(\delta) = \text{sin}[\Delta\theta]$, if ($\alpha,\beta,\gamma$) are constant parameters. This situation corresponds to spinor BECs with trivial spinor parts (for instance, in the ferromagnetic case, $\boldsymbol{\zeta}=(1,0,0)^T$ or $(0,0, 1)^T$), representing an effective scalar BEC in a rotated reference frame.

\subsection{Polar State}

The \textit{polar} states are generated by $\boldsymbol{\zeta}_0^{\text{polar}} = (0,1,0)^T$. Such a family of states is unique because they are closely related to scalar BECs (given that $f_0=0$ and $\vec{F}=0$), while their symmetrical and topological properties might be nontrivial \cite{Kawaguchi_Ueda}. Using (\ref{matriz}), one finds the spinor part of the polar states:
\begin{equation}\label{polar}
    \boldsymbol{\zeta}_{\text{polar}} = \begin{pmatrix}-\frac{e^{-i\alpha}}{\sqrt{2}}\text{sin}\beta\\ \text{cos}\beta \\ \frac{e^{i\alpha}}{\sqrt{2}}\text{sin}\beta\end{pmatrix}.
\end{equation}
{Applying}  this expression in (\ref{B}), we obtain $C_{\text{polar}}(\delta)$:
\begin{eqnarray}
    C_{\text{polar}}(\delta) &=& \frac{1}{2}\text{sin}\beta(0)\text{sin}\beta(\delta)\text{sin}[\Delta\theta-\Delta\alpha] + \text{cos}\beta(0)\text{cos}\beta(\delta)\text{sin}[\Delta\theta] \nonumber \\  
    &+& \frac{1}{2}\text{sin}\beta(0)\text{sin}\beta(\delta)\text{sin}[\Delta\theta+\Delta\alpha]. 
\end{eqnarray}

{With} a little algebraic work, we find that $C_{\text{polar}}(\delta)\propto \text{sin}[\Delta\theta]$ in general. In this sense, we might interpret such a relation as a signature of the \textit{partial analogy} between \textit{polar states} and \textit{scalar} BECs, because the current density $J_{\varphi}(\delta)$ can be written in the following way:
\begin{equation}
    J_{\varphi}(\delta) = J_{\text{max}}(\alpha,\beta)\,\text{sin}[\Delta\theta],
\end{equation}
where the parameters $(\alpha, \beta)$ modulate the maximum value of $J_{\varphi}(\delta)$.

\section{Landau Critical Velocity, Lower Bound for $R$, and Experimental Protocol} \label{S6}

This section discusses the effects of the Landau critical velocity on the parameters of our model and shows that it defines a lower bound for $R$. We also introduce a simple experimental scheme to test the theory, proposing an upper limit to how fast the barrier height, $U_0(t)$, can be turned on. In addition, we discuss some relevant considerations and practical suggestions for the experiments, particularly related to time-lapsed measurements of the BEC density distribution, using specialized imaging techniques, from which the analysis of the time evolution would lead to $C(\delta)$.

The Landau critical velocity is an upper limit for the velocity of particles in a superfluid, before the appearance of dissipation (viscosity) \cite{pethick,Pitaevskii2016,Bogoliubov,velocidadeLandau}. It is set by the energy gap between the ground state and the lowest elementary excitation leading to dissipation in the fluid. In BECs, this critical velocity is typically in the same order of magnitude as the \textit{sound velocity}, being identical to it for a weakly-interacting homogeneous (uniform) scalar BEC. According to the Bogoliubov theory for spin-1 spinor BECs \cite{Kawaguchi_Ueda}, similarly to the scalar case, this velocity $v_c$ is
\begin{equation}
    v_c \sim \sqrt{\frac{gn}{M}},
\end{equation}
where here $g$ is usually a linear combination of the constants $c_0$ and $c_1$ from Equation~(\ref{GPE}), and it depends on the broken-symmetry state of the condensate. For example,  in a ferromagnetic state,  according to Sec. 5.2.1 in Ref. \cite{Kawaguchi_Ueda}, $g=c_0+c_1$. 

Because $J_{\varphi}(\varphi)$ is an odd (anti-symmetrical) function in our formulation, from symmetry arguments alone, we expect $J_{\varphi}(0)=J_{\varphi}(\pi)=J_{\varphi}(-\pi)=0$, with the maximum amplitudes symmetrically occurring in the interval $|\varphi|\in [0,\pi]$. Therefore, assuming that the dominant contribution to the current occurs at $\varphi=\pm \delta$, because we are interested in the superfluid regime, we impose $|\vec{v}(\delta)|<v_c(\delta) = \sqrt{gn(\delta)/M}$, where $n(\delta)$ is the total (local) density at $\varphi=\pm \delta$. 
Moreover, we use Equation (\ref{campo de velocidades}) to estimate $|\vec{v}(\delta)|$:
\begin{equation}
    |\vec{v}(\delta)| \sim \frac{\hbar}{MR}|\partial_{\varphi}\theta(\delta)-i\boldsymbol{\zeta}^{\dagger}(\delta)\partial_{\varphi}\boldsymbol{\zeta}(\delta)| \sim \frac{\hbar}{MR} \frac{1}{\delta}.
\end{equation}

Considering $\delta$ is a given angle (i.e., we chose $\delta$ in an experiment), we can estimate a lower bound for $R$, such that the condensate is in the superfluid regime:
\begin{equation}
    R>\sqrt{2} \frac{\hbar}{\sqrt{2Mgn(\delta)}} \frac{1}{\delta} = \sqrt{2} \frac{\xi}{\delta}; \quad \text{with }\xi \equiv \sqrt{\hbar^2/(2Mgn(\delta))}.
\end{equation}

Or, more intuitively:
\begin{equation}
    R\gg \frac{\text{(Healing Length)}}{\delta}
\end{equation}
if one wants the system to be far from the critical velocity.

As indicated previously in Equation (\ref{continuityequation}), the currents $J_{\varphi}(\pm\delta)$ are helpful to model the rate of change in the number of atoms $N_B$ leaving the region inside the barrier. Conversely, measuring this rate of change, i.e., monitoring $N_B(t)$, is a way to measure $C(\delta)$ as a function of time. In principle, one could use non-destructive imaging techniques \cite{Andrews1996,Ramanathan2012} to take multiple snapshots of the condensate density distribution at different times. Depending on the experimental parameters, such as the number of atoms and the radius $R$, the cross-section $A_c$ could be so small that the optical density of the atoms would make dispersive (phase contrast) imaging \cite{Andrews1996} difficult, but this is exactly the conditions for which partial-transfer absorption imaging (PTAI) \cite{Ramanathan2012} was developed. Therefore, in principle, one can make these measurements even in the deep quasi-1D limit (i.e., extending the radius $R$ as necessary to fulfill the condition in Section \ref{S4}).

According to Section \ref{S5}, different types of spinor BECs (ferromagnetic, polar, and scalar, with equivalent initial conditions for the density and velocity) should respond differently to the application of the same potential barrier $U_B(\varphi)$. The details will depend on the specific case, as well as the external magnetic field. Therefore, further theoretical, numerical, and experimental studies are necessary, but they may unveil new ways of sensing the symmetry of spinor BEC states, solely using  measurements of $C(\delta)$.

As a simple example, we can imagine an experimental protocol starting initially with the barrier turned off and a homogeneous condensate in equilibrium ($n(0) = n(\delta)$ and $\partial_t n(\varphi)=0$). At $t=0$, the barrier is turned on and ramped up following a sufficiently fast time protocol $U_0= U_0(t)$. The barrier causes a perturbation on the densities (and also on $N_B$ and $C(\delta)$), and, after a time $t=\tau$, the system reaches a new equilibrium ($\partial_t N_B = 0$ and $C(\delta)=0$, $\forall t>\tau$). For scalar BECs, the equilibrium is described by $C(\delta) = \text{sin}(\Delta\theta)=0$, with $\Delta\theta = $ constant, which (as expected) simply implies in:
\begin{equation}
    \mu = \left(\frac{\delta E}{\delta n}\right)_{\varphi = \delta} = \left(\frac{\delta E}{\delta n}\right)_{\varphi = 0}.
\end{equation}

Finally, we briefly show here that the time protocol $U_0(t)$ must always have an upper limit that depends on the density's perturbation and $C(\delta)$ at any given time so that the condensate does not lose its superfluid properties. This upper limit is found using Equation~(\ref{RAREcurrent}) to estimate $|\vec{v}(\delta)|$ and it corresponds to
\begin{equation}
    U_0(t) < v_c(\delta,t) \frac{\hbar}{R} \sqrt{\frac{n(\delta,t)}{n(0,t)}}\frac{1}{|C(\delta,t)|}.
\end{equation}

\newpage
\section{Conclusions}

In this manuscript, we used the parallel between superconductors and superfluids to present a curious new effect in a superfluid, similar to the current oscillations in the Josephson effect, but it happens without current flowing through the barrier. The result was derived for BECs in toroidal traps with a thin finite repulsive barrier, and it works both for scalar and spinor BECs.

Using a thin barrier approximation, in Equations (\ref{eq8}) and (\ref{eq9}), and assuming a defined parity for the order parameter, we derived analytical expressions for the current density. We showed that the current at the edges of the barrier oscillates in a similar fashion to the current flowing through a Josephson junction (\ref{JosephsonCurrent}), but in our case, it happens without a net circulation of the superfluid current. 
We also showed how the nontrivial symmetry properties of spinor BECs could generate other current structures, beyond the typical $C(\delta) = \text{sin}[\Delta\theta]$ case, indicating that such current is sensitive to these properties, suggesting that it could be used to probe spinor symmetry or, perhaps,  provide precision measurements related to the superfluid flow.

Throughout this work, we assumed the parity of $\boldsymbol{\psi}$ was defined, which is a strong mathematical imposition to guarantee that Equation (\ref{RAREcurrent}) is a reasonable approximation for the current density. Moreover, this imposition implies that the circulation of the condensates considered here is always zero, excluding several $(\theta,\alpha,\beta,\gamma)$ configurations with nontrivial symmetry and topology. However, as we discussed \textcolor{black}{in Section \ref{S4}}, ensuring that the last term in (\ref{DeltaJ_completa}) is dominant should keep the structure of $\Delta J(\delta)$ unchanged, regardless of the parity of $\boldsymbol{\psi}$. Therefore, it might be possible to derive a similar expression (\ref{RAREcurrent}) for BECs with nonzero circulation, such as the ones shown in experiments \cite{Ramanathan2011,Wright2013,Eckel2014,Beattie2013,Moulder2012}, and for spinor BECs with nontrivial structure or topology \cite{Kawaguchi_Ueda,Leanhardt2002,Leanhardt2003,Kumakura2006,Wright2008,Wright2009,Leslie2009,JDiaz2009a,JDiaz2009b,Mateo2015}.

Here, we also neglected the effects of fluctuations, either classical or quantum, which can be relevant in the quasi-1D limit \cite{Petrov2001,AlKhawaja2003,Gerbier2004}. However, as we have shown previously in Ref. \cite{Mathey2010}, one can always play with the number of atoms and the aspect ratio of the trapping potential to place oneself within the best range of parameters for the experiments. Exercising this ability deliberately allows one to control the influence of thermal phase fluctuations at finite temperatures. Therefore, interesting future directions would be to explore the effects of finite temperature and thermal fluctuations, and the non-equilibrium effects caused by quickly turning on the repulsive barrier in our proposed experimental scheme.
In Section \ref{S6}, we discussed some considerations for an idealized protocol, but the general spinor case is more complex and deserves a detailed analysis. For instance, our simple estimate for the Landau critical velocity does not take into account other relevant mechanisms of decay \cite{Piazza2013,Moulder2012,Beattie2013}, especially at finite temperatures \cite{Mathey2014}.

Therefore, further numerical, experimental, and theoretical studies are necessary and may improve, expand, and, perhaps, help to classify (maybe in symmetry terms) the formulation of this intriguing Josephson-like effect in toroidal spinor Bose--Einstein~condensates.

\begin{acknowledgments}
\noindent We acknowledge the financial support provided by FAPESP (Fundação de Amparo à Pesquisa do Estado de São Paulo), the São Paulo Research Foundation, under Grant no. 2013/07276-1 and 2019/27471-0, and also by CAPES Grant nos. 88887.338144/2019-00 and 88887.616990/2021-00.
\end{acknowledgments}

\end{document}